\documentclass[conference,compsoc]{IEEEtran}
\ifCLASSOPTIONcompsoc
  \usepackage[nocompress]{cite}
\else
  \usepackage{cite}
\fi
\ifCLASSINFOpdf
\else
\fi
\usepackage{tikz}
\usepackage{amsmath}
\usepackage{graphicx}
\usepackage{booktabs}
\usepackage{multirow}
\usepackage{float}
\usepackage{hyperref}
\usepackage{amsfonts}
\usepackage{tcolorbox}

\begin{document}

%
\title{Know Your Neighborhood: General and Zero-Shot Capable Binary Function Search Powered by Call Graphlets}

\author{\IEEEauthorblockN{Josh Collyer, Tim Watson and Iain Phillips}
\IEEEauthorblockA{Loughborough University \\
Epinal Way \\ 
Loughborough \\
LE11 3TU} }


%


\maketitle

\begin{abstract}
Binary code similarity detection is an important problem with applications in
areas such as malware analysis, vulnerability research and license violation detection.
This paper proposes a novel graph neural network architecture combined with a
novel graph data representation called \textit{call graphlets}. A call graphlet encodes
the neighborhood around each function in a binary executable, capturing the
local and global context through a series of statistical features. A specialized
graph neural network model operates on this graph
representation, learning to map it to a feature vector that encodes semantic
binary code similarities using deep-metric learning.

The proposed approach is evaluated across five distinct datasets covering
different architectures, compiler tool chains, and optimization levels.
Experimental results show that the combination of call graphlets and the
novel graph neural network architecture achieves comparable or state-of-the-art performance
compared to baseline techniques across cross- architecture, mono-architecture
and zero shot tasks. In addition, our proposed approach also performs well when
evaluated against an out-of-domain function inlining task. The work
provides a general and effective graph neural network-based solution for
conducting binary code similarity detection.
\end{abstract}


%
\IEEEpeerreviewmaketitle

\section{Introduction}
Binary code similarity detection (BCSD) aims to find compiled functions that are
similar to a given compiled query function. It has a wide range of applications,
from identifying vulnerabilities in firmware to analyzing code reuse across
malware variants. BCSD can enhance security assessments of closed-source
software by identifying the components used, such as libraries
embedded within statically linked executables and firmware binaries.

Implementing effective BCSD approaches is challenging due to the diversity of
architectures, compiler tool chains, and optimization levels used to build
software. These factors can lead to significant variations in the binary
representations of the same source code, making it challenging to distinguish
between similar and dissimilar functions. While optimization level has the
biggest impact on these variations, recent research shows that call graph
features are most resilient to these transformations, presenting an opportunity
to develop improved, general BCSD approaches \cite{kim2022revisiting}.

Existing approaches predominantly focus on function-level artifacts like control
flow graphs, disassembly, or intermediate representations, neglecting the
potential of call graph features. Additionally, some approaches rely on the
assumption of dynamically linked standard library functions such as those from
\textit{libc} are easily recoverable. This does not hold for embedded devices
and malware where static linking is common, restricting their applicability
across different software ecosystems.

Furthermore, the evaluation methodologies employed in binary function search
often lack rigor. Existing approaches predominantly rely on ``in-domain''
evaluation data, utilizing subsets of functions from the training binaries
themselves. This practice introduces the risk of data leakage, whereby the model 
is evaluated on the same data in which it was trained leading to the potential of
reporting inflated performance metrics. More crucially, it fails to assess the
true generalization capability of the approach when faced with previously unseen
functions, a scenario referred to as ``out-of-domain'' data. Comprehensive
evaluation necessitates testing on entirely separate datasets to obtain an
accurate understanding of real-world performance and generalizability. This is
an area we tackle head on with evaluation across five distinct datasets.

This paper proposes a novel approach that aims to exploit the additional
information available when including call graph features and addresses the
limitations of existing methods. It employs a comprehensive evaluation
methodology focusing on out-of-domain assessments across four
large datasets, to ensure robust and generalizable findings. The proposed
approach does not rely on the identification of standard library functionality,
making it applicable across a broader range of binary types and scenarios.

\subsection{Key Contributions}

Within this work, we propose a general function search approach utilizing
function level call graphlets inspired by the results presented in TikNib
\cite{kim2022revisiting} and PSS \cite{benoit2023scalable}. Our contributions
are as follows:
\begin{enumerate}
  \item We introduce a novel call graph level function format we name call
  graphlets. This format incorporates a functions callers as well as callees to
  provide a graphlet neighborhood which provides structure alongside a set of
  simple function level features as node attributes.
  \item We present a simple, yet powerful, Graph Neural Network (GNN) trained
  using deep-metric learning capable of performing well in and out of
  domain across a number of tasks. 
  \item We present a robust evaluation of our approach using a range of research dataset 
  benchmarks and compare against both sequence and graph based state-of-the-art approaches.
  \item We conduct an ablation study to investigate how our graph neural network
  (GNN) design choices contribute to our performance.
\end{enumerate}

The data, models and associated experimental code is available: \url{https://github.com/br0kej/kyn}

\subsection{Paper Overview}

The remainder of this paper is organized as follows: We begin by reviewing the
relevant prior work to establish a foundation for our contributions. Next, we
present a detailed description of the methodology used in this study. This
includes explaining how a call graphlet is constructed, the features used for
nodes and edges, the Graph Neural Network (GNN) architectural choices, and the
datasets employed. Following this, we present a comprehensive section on the
experiments conducted to evaluate our approach. We start with an evaluation in a
cross-architecture context, followed by experiments evaluating the model's
performance in a mono-architecture setting (x86-64 only), a cross-architecture
with function inlining setting, a zero-shot vulnerability search setting, and
then conclude with an ablation study. Finally, we conclude the paper by
exploring some fundamental limitations of our approaches and providing our
concluding remarks.

\section{Related Work}
The problem of identifying similar functions across different instruction set
architectures has been tackled using various methods in the existing literature.
These methods can generally be classified into two main categories:
sequence-based approaches or graph-based approaches. In this section, we start
by briefly reviewing the sequence-based approaches. Then, we provide a more
detailed review of the graph-based literature. Finally, we briefly introduce a
relevant alternative method and before then highlighting the limitations of
previous works in this area.

\subsection{Sequence-Based Approaches}

Sequence-based approaches use either assembly, pseudocode or intermediate
representations extracted from binaries for each function to generate
representations suitable for similarity comparison. These sorts of approaches
have been applied to individual assembly instructions\cite{li2021palmtree},
sequences of assembly instructions\cite{wang2022jtrans}, sequences of
intermediate representations\cite{collyer2023faser} and even execution sequences
\cite{pei2020trex}. 

All approaches share the common objective of seeking to learn useful
representations of input by adapting approaches developed in the natural
language processing (NLP) literature. This is typically done via a two stage
process of pre-training and then fine-tuning. Researchers have used a range of
different pre-training tasks ranging from standard approaches such as Masked
Language Modelling (MLM) to domain specific tasks such as jump target prediction
\cite{wang2022jtrans} and instruction data flow relationship prediction
\cite{li2021palmtree}. Once pre-trained, the models are then fine-tuned for a
given down stream task such as binary code similarity detection (BCSD) or N-day
vulnerability detection typically using contrastive methods such as deep-metric
learning.

\subsection{Graph Based Approaches}

All of these approaches however miss an opportunity to use broader information
related to a single function such as type information and calling relationships
and instead focus on data contained within a single function boundary. Graph
based approaches on the other hand, look to exploit the inherent structure
within compiled binaries such as call graphs and control flow graphs (CFG).
These data structures are then paired with graph neural network approaches to
create learned representations that perform well at downstream tasks such as BCSD
or N-day vulnerability detection.

XBA \cite{kim2022improving} is a graph neural networks approach which works on a
basic block level for a given function as well as incorporating key external
information such as string literals and external function calls. This novel
representation is then fed into a GCN-based GNN to generate cross-architecture
and cross-platform embeddings. PDM \cite{pan2022position} is another example of
a graph neural network which instead works on a combined control flow and data
flow representation named \textit{ACFG+}. This representation is then used
alongside a Capsule GNN to generate embeddings. 

CFG2VEC \cite{yu2023cfg2vec} is an example of a more complicated approach to
ours and was used as inspiration for our approach. CFG2VEC is a hierarchical
method whereby a single input is the call graph of the binary of interest and each node is a
P-code Based control flow graph. The P-Code based CFG representation is then fed
into a graph neural network to create node-embeddings for the higher level call
graph. These node embeddings are then subsequently used to support binary code
similarity detection tasks. PSS \cite{benoit2023scalable} propose an interesting
graph spectra approach which operates primarily on the call graphs structure.
This approach is used for whole binary identification so can be considered
tangential to the other papers within this section but has been included
primarily because it was also a large the main inspiration for this work.
HermesSim \cite{he2024code} also used a P-Code representation of a target
function but instead creates a novel graph representation where nodes are P-code
operations or memory locations and the edges are relationships such as read and
writes. This is again fed into a GNN approach to create function level
representations that can be used for BCSD tasks. HermesSim can be considered the
current state of the art in this area having reported impressive results
recently and is the primary baseline we use within this work.

\subsection{Alternative Methods}

In addition to sequence, graph and combinatorial methods, some researchers have
sought to adopt alternative methods. BinFinder \cite{qasem2023binary} is an
example of this were instead of using sequences, a series of statistical count
features that are calculated from the VEX IR lifted representation of a function
alongside counts of function calls such as those to \textit{glibc}. This
approach has one major drawback and that is the reliance on glibc functions
being recoverable and external. This limits its usage in cases
such as statically linked binaries or blob firmware. The approach however does
perform very well and is one our key baselines.

\subsection{Limitations of Previous Work}

There are several limitations of the works mentioned above. Firstly, several of
the approaches \cite{qasem2023binary,kim2022improving} rely on recoverable
function names to build fundamental parts of their feature space. This is
error-prone and something that may not always be possible in all settings such
as when you are faced with obfuscated or statically linked binaries. Secondly,
most existing approaches
\cite{pan2022position,li2021palmtree,collyer2023faser,wang2022jtrans,pei2020trex}
analyze individual functions in isolation without considering the broader
relationships and interactions between functions within the software binary.
This potentially overlooks valuable information that is inherent in the overall
structure and interconnections present in the binary.

Thirdly, when approaches utilize external information like global call graphs
\cite{yu2023cfg2vec,benoit2023scalable}, they encounter two primary limitations.
First, a global call graph is a coarse representation, and its structure, and
consequently its effectiveness, can be significantly affected by obfuscation
techniques and function inlining. Secondly, errors in accurately recovering the
correct calling relationships may propagate throughout the entire graph, even
though these errors may only impact a subset of the functions within the binary
executable, the effect may still be significant.

\section{Methodology}

\subsection{Call Graphlet Neighborhoods}

\begin{figure}[h]
  \begin{center}
    \includegraphics[scale=0.5]{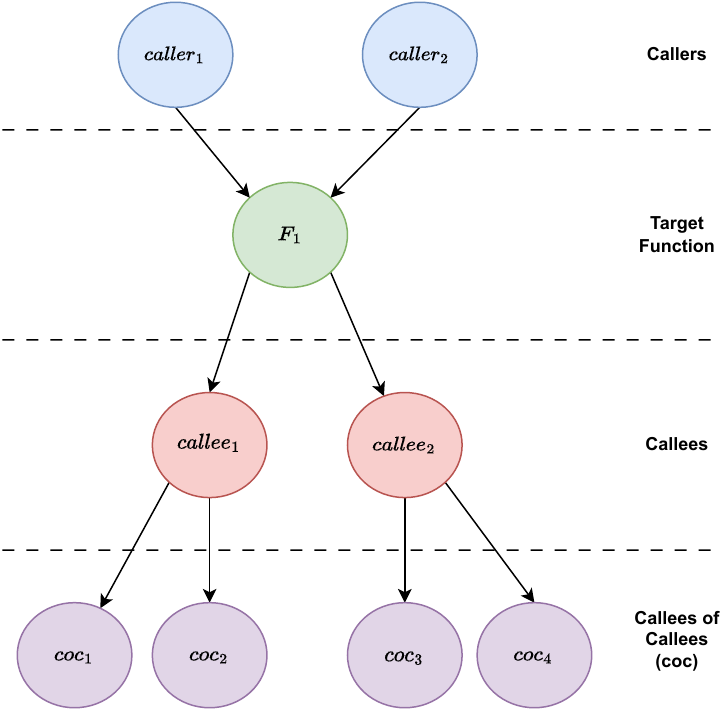}
  \end{center}
  \caption{Call Graphlet Example Data Structure}
  \label{fig:example_graphlet}
\end{figure}

To overcome these limitations, our work utilizes a novel data format we call
\textit{call graphlets} which is shown in Figure~\ref{fig:example_graphlet}. This approach was inspired by the work described in
PSS \cite{benoit2023scalable} but instead of operating with binary-level call
graphs, we opt to work with function-level ones which provide a balance of local
and global information. 

A call graphlet is a weighted directed graph made up of a tuple $G = (V,E,W)$
where $V$ represents a set of functions, $E$ represents a set of calling
relationship in the format of $ (x,y) \; | \; x \neq y$ and $W$ represent the
weights for a given edge. A target function's call graphlet neighborhood
includes a node for the target function itself, any functions that call the
target function (callers), any functions the target function calls (callees) and
any functions that are called by the target functions callees (callees of
callees). We excluded callers of callers from our call graphlets to prevent
creating excessively large call graphlets, particularly for widely used
functions like utilities and standard libraries, and because we felt that callee
relationships better characterize a function. The weights for all edges
correspond to there respective edge betweenness which is calculated at a call
graphlet level rather than a global call graph level.

The reasons for working at a call-graphlet level are twofold. Firstly, recent
work suggests that features derived from the call graph are the most robust to
changes in compilation tool chains \cite{kim2022revisiting}. We however viewed
using global call graphs as too coarse and instead formulated a more
fine-grained data representation with the aim of developing a high performing
function-level BCSD approach. Secondly, we wanted to create a data
representation that took the calling behaviour between a function, its callers
and callees into account. We view this relationship as something that will
likely be similar within software compiled from the same source code regardless
of compilation variations. This provides a means to focus on a higher level
representation rather than solely focusing on function level differences which
can be effected by not only toolchain choice but also changes such as patches or
refactoring.

\subsection{Graph Features} 

\subsubsection{Node Features}

Each of the functions within a given call graphlet neighborhood have six
straightforward features derived from function level metadata recovered during
the disassembly process. Features such as in-degree (number of callers),
out-degree (number of callees) and total number of edges provide global
information whilst features such as total number of instructions, number of
arguments and number of local variables provide local information. The
hypothesis is that this combination of local and global level information
provides the model the ability to discern between a range of different
functions. For example, a complex function that is infrequently called may have
a low in-degree but a high number of instructions, out-degree and number of
local variables whereas a wrapper function for a commonly used library function
such as \textit{memcpy} may have a low instruction count and out-degree but a
very high in-degree. The full feature list can be found in Table~\ref{table:node_features}.

\begin{table}[ht]
  \begin{tabular}{lp{5.5cm}}
  \toprule
  \textbf{Feature Name}           & \textbf{Description} \\ \midrule Num
  Instructions & The number of instruction in the function \\
  Num Edges                  & The sum of in-degree and out-degree edges \\
  Total Indegree               & The number of in-degree edges. Denotes callers
  \\ 
  Total Outdegree              & The number of out-degree edges. Denotes callees
  \\
  Num Locals                & The number of local variables \\
  Num Args               & Number of arguments from the recovered function
  prototype \\ \midrule
  \end{tabular}
  \caption{Function/Node Level Features}
  \label{table:node_features}
\end{table}

\subsubsection{Edge Weights}

Alongside the node-level function metadata, we also generate a weight for each
edge within the call graphlet using edge betweenness centrality.

The edge betweenness for edge \(e\) is calculated as follows:
\begin{equation}
  C_b(e) = \sum_{s,t \in V} \frac{\sigma (s,t|e)}{\sigma (s,t)}
\end{equation}
\label{eq:edge_betweenness}

Where \(V\) is the set of nodes, \(\sigma (s,t)\) is the number of shortest
paths between the nodes \(s\) and \(t\) and \(\sigma (s,t|e)\) is the number of
shortest paths that pass through \(e\).

The edge betweenness is calculated at a call graphlet level instead of a global
call graph level because we want to quantify the control each edge exerts over
information flow within the call graphlet. Edges with high betweenness
centrality can be viewed as critical pathways or bottlenecks. In our particular
case, we viewed edge betweenness centrality as a useful addition because it
provides a means of identifying functions that exhibit high call fan-out (i.e.,
a large number of callees) and participate in critical calling relationships
with ancestor functions in the call hierarchy. This additional information is
hypothesized to improve the model's performance by capturing the interplay
between a function's calling behavior and its position within call graphlet's
overall structure.

\subsection{Datasets}

\subsubsection{Training Dataset}

The training dataset used within this paper is the \textit{Dataset-1} released
as part of the work of \cite{marcelli2022machine}. This dataset contains seven
popular open source projects: ClamAV, Curl, Nmap, Openssl, Unrar, z3 and zlib.
Each of these are compiled for ARM32, ARM64, MIPS32, MIPS64, x86 and x86-64
using 4 different versions of Clang and GCC alongside 5 different optimization
levels. Each project has 24 unique configurations. The same sized train, test
and validation splits were created as described in \cite{marcelli2022machine}
and \cite{he2024code} to support robust comparison.

It is worth noting that the processed call-graphlet dataset created from
\textit{Dataset-1} using \textit{bin2ml}\cite{bin2ml} is significantly larger than what is
reported in \cite{marcelli2022machine} and \cite{he2024code}. This is because
our data representation allows us to relax the filtering requirements such as
number of instruction or basic blocks within a function. Since our method uses a
function's neighborhood, we can include smaller functions as they will be
augmented by their neighbors who may not fit the filtering requirements.

Our data representation also likely results in more varied examples of each
function because, due to architectural, compiler or optimization differences,
the variations present within a function's neighborhood are included as well.
This is especially interesting because even when the node features of a function
are duplicates, each call graphlet may be different and therefore, continue to
be included within our dataset. With this being said, in order to facilitate
robust and fair comparison, we firstly subset the data into the same
binary-level splits described in \cite{marcelli2022machine} and then use random
sampling to create an identically sized sample from within the binaries
previously subset. This methodology is used to generate the both the training
and tests sets for \textit{Dataset-1}.

\subsubsection{Evaluation Datasets}

In order to robustly evaluate our approach, we use five open source benchmark
datasets and focus exclusively on out-of-domain evaluation. The reason for
focusing exclusively on out-of-domain performance is that it provides
experimental evidence that our approach performs well when faced with a
collection of binaries it has not seen before as well as being the most
comparable to a real-world setting.

Firstly, we use the test set from {Dataset-1}. This dataset is a collection of
binaries which encompasses the Z3 and Nmap projects, all of which are not
present within the train dataset. This dataset contains 522003 functions which
is identical to the size used within \cite{he2024code} and
\cite{marcelli2022machine}.

Secondly, we use \textit{Dataset-2} also presented in
\cite{marcelli2022machine}. This dataset is a subset of the evaluation dataset
used within \cite{pei2020trex} and ensures there is no intersection between the
software packages contained within \textit{Dataset-1}. \textit{Dataset-2}
includes the following 10 libraries: \texttt{Binutils},\texttt{coreutils},
\texttt{diffutils}, \texttt{findutils}, \texttt{GMP}, \texttt{ImageMagick},
\texttt{Libmicrohttpd}, \texttt{LibTomCrypt}, \texttt{PuTTy} and
\texttt{SQLite}. Due to issues processing \textit{ar} archives with our data
processing tool, \texttt{Libmicrohttpd} and \texttt{LibTomCrypt} were omitted
from our evaluation dataset. Both of these binaries are however included within
the later datasets. The binaries within this dataset include versions compiled
for x86, x64, ARM32, ARM64, MIPS32 and MIPS64 across four optimization levels
(O0, O1, O2, O3) using the \texttt{GCC-7.5} compiler. This dataset shares the
same overall configuration in terms of architectures, optimization levels and
compiler version but has unique binaries within it.

Thirdly, we use \textit{BinKit NoInline} dataset presented within
\cite{kim2022revisiting}. This dataset contains 51 unique packages/libraries
across 8 architectures (x86, x64, ARM32, ARM64, MIPS32, MIPS64, MIPS32EB and
MIPS64EB), 9 different versions of compiler (five unique versions of GCC, four
unique versions of Clang) as well as five different optimization levels (O0, O1,
O2, O3, Os). All the binaries have also been compiled with function inlining
turned off. This results in a total of 67,680 binaries. This dataset does share
some common software projects to \textit{Dataset-2} but does not share any with
the training dataset. The dataset introduces two previously unseen architecture
variants, additional compilers as well as an additional optimization level. 

Fourthly, we use the \textit{BinKit Normal} dataset also presented within
\cite{kim2022revisiting}. This is an identical dataset to the \textit{Binkit
NoInline} dataset described above but with one fundamental difference - The
compilers are able to inline functions. Function inlining has been referenced as
a significant weakness across a number of binary function similarity approaches
\cite{wang2022enhancing, gao2019semantic} and therefore this dataset allows us
to tackle this area head on and generate empirical results of how well our
approach fairs when faced with function inlining.

Finally, we use the \textit{Dataset-Vulnerability} dataset presented as part of
\cite{marcelli2022machine}. This dataset is a collection of OpenSSL libraries
compiled for a range of architectures alongside two examples of the identical
library but instead taken from a real device, one which is \texttt{arm32} and
another which is \texttt{mips32}. This forms the basis for a small scale N-day
vulnerability benchmark.

\subsubsection{Data Preparation}

A uniform data preparation pipeline was implemented to process all datasets
outlined above except the vulnerability benchmark. This process can be broken
down into five key stages: extraction, fusion, deduplication, augmentation and
sampling.

\noindent\textbf{Extraction:} The initial stage involved extracting call graphs
and function metadata for each binary within the datasets. This was achieved by
utilizing the \textit{bin2ml}\cite{bin2ml} tool which uses \textit{radare2}\cite{radare2} as the underlying
software reverse engineering tool to load, process and output the required data.

\noindent\textbf{Fusion:} Following extraction, the extracted call graph and
function metadata were combined. This process resulted in the creation of
function-level call graphlets. Each call graphlet encapsulated the functions
local call graph structure whereby each node's features was it's corresponding
function's metadata. Each call graphlet was output as a \textit{Networkx}
\cite{hagberg2008exploring} compatible JSON object to support loading
into our chosen graph neural network framework, \textit{PyTorch Geometric}
\cite{fey2019pygeometric}.

\noindent\textbf{Deduplication:} The next stage focused on eliminating redundant
call graphlets within the datasets. The methodology employed here was inspired
by previous related work \cite{marcelli2022machine,collyer2023faser,he2024code}.
The deduplication process is outlined as follows:

\begin{enumerate}
  \item Initially call graphlets are grouped based on the software binary from
  which they originated. This binary-specific grouping facilitates a more
  efficient deduplication process. 
  \item  Subsequently, each call graphlet within a group is hashed. The resulting hashes are then stored within a hash
  set. This hash set serves as a mechanism to track unique call graphlets
  encountered for each binary.
  \item In the event that a duplicate hash is identified, the call graphlet is
  removed from the dataset. This ensures the presence of a single example for
  each unique call graphlet.
\end{enumerate}

All the remaining unique call graphlets are saved and form the basis of the
given dataset used to train or evaluate our proposed approach.

\noindent\textbf{Augmentation:} Once deduplication was complete, each remaining
graph was loaded into \textit{Networkx} and the edge-betweenness centrality
calculated and then appended to the graph as an edge weight.

\noindent\textbf{Sampling:} The last stage focused on sampling the data. This was 
only necessary for \textit{BinkitNormal} and \textit{BinkitNoInline}. Due to the 
scale of the datastets once processed, it was not feasible to evaluate the model of 
all functions due to hardware limitations. To overcome this, we used a random sampling
strategy to create test datasets. Each sample contained a total of 1 million
functions. This subset was created once for each source dataset and then saved
for re-use. Both of these splits are provided as part of the release of this
research. 

\subsection{Model}

\begin{figure*}[h]
  \begin{center}
    \includegraphics[scale=0.4]{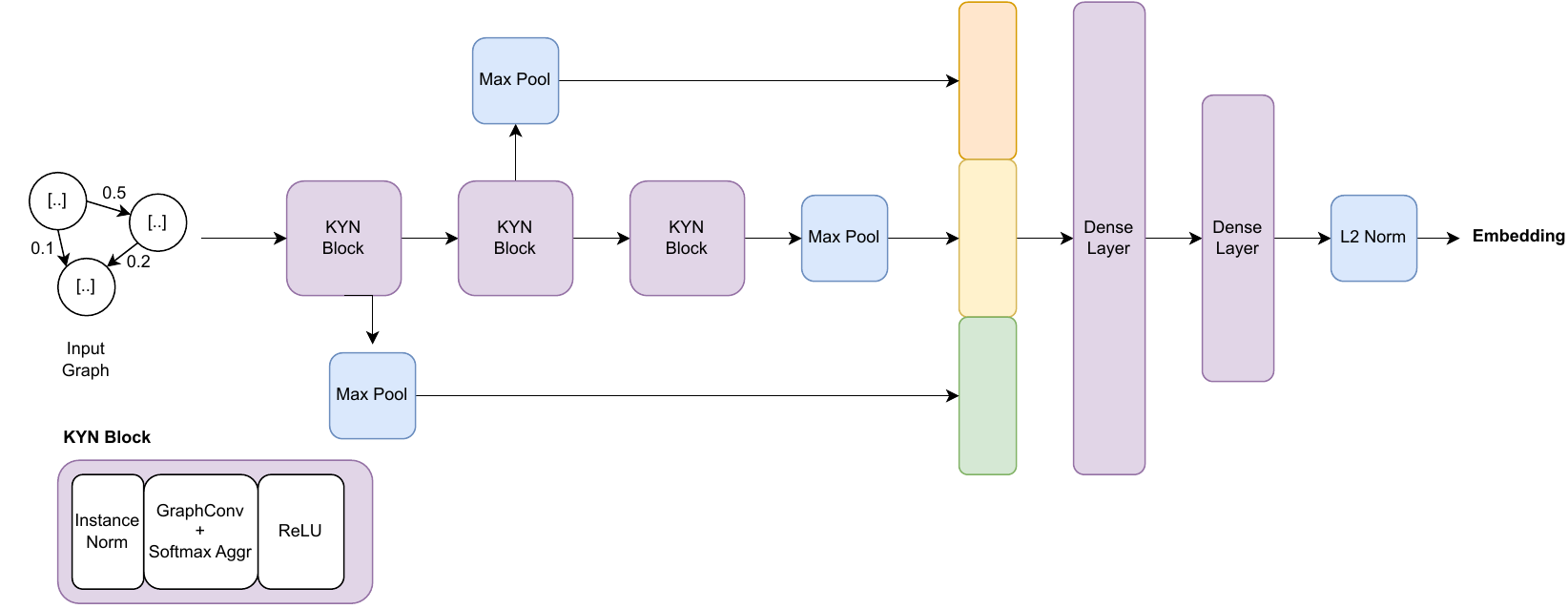}
  \end{center}
  \caption{KYN Model Architecture}
  \label{fig:model_arch}
\end{figure*}  

The model used within our approach is a 3-layer  
graph neural network incorporating several GNN components to create a novel GNN
architecture. A diagram visually representing the model can be seen in Figure~\ref{fig:model_arch}

\noindent\textbf{GraphConv Message Passing GNN}: The graph convolution operator
used within this study is the operator proposed in \cite{morris2019weisfeiler}
and is shown in Equation~\ref{eq:graphconv}. It computes the updated feature
representation \(x_i'\) for a node \(i\) in a graph by combining two components:
1) a linear transformation of the node's own input features \(W_1x_i\), and 2) a
weighted sum of the features of its neighboring nodes \(\sum_{j \in N(i)}
e_{j,i} \cdot x_j\), where the neighbors' features are first weighted by the
strength of their connections \(e_{j,i}\) and then linearly transformed by \(W_2\).
The resulting output feature vector \(x_i'\) captures both the node's intrinsic
features and the information from its local neighborhood. This operator has been
shown to be able to capture higher-order graph structures and outperform a range
of standard GNN and graph kernel approaches. This is due to approach considering
structures within the graphs at varying scales. The properties of
this operator suit our call graphlet data representation and help
leverage the richness of a functions calling relationships.

\begin{equation}
  x_i \prime = W_1x_i + W_2 \sum_{j \in N(i)} e_{j,i} \cdot x_j
\end{equation}
\label{eq:graphconv}

\noindent\textbf{Instance Normalization}: We adopt Instance Normalization (IN)
\cite{ulyanov2016instance} as the graph normalization technique omitting the
learnable scale and shift parameters \( \gamma \) and \( \beta \), as described
in Equation~\ref{eq:instance}. IN is applied to the node features, where
\(x_i\prime\) represents the normalized node feature for node \(i\). It centers
the features by subtracting the mean \(E[x]\) across all nodes within a given
graph, scales them by dividing by the square root of the variance \(var[x]\)
plus a small constant \(\epsilon\) (set to \(1e-05\)). This normalization
technique aims to stabilize the training process by reducing the internal
covariate shift, ensuring consistent distribution of node features across the
layers of the graph neural network.

\begin{equation}
  x_i{\prime} = \frac{x - \mathit{E}[x]}{\sqrt{\mathit{var}[x] + \epsilon}}
  \label{eq:instance}
\end{equation}

The selection of
architecture, compiler, and optimization combinations can significantly
influence a function's structure and form therefore introducing a large variance
in our simple, node level statistical features. IN exhibits an interesting
property wherein it performs normalization on each function independently at the
graph level. This means that variations among features within our examples are
handled separately, which is crucial due to substantial variations
across multiple features, such as the number of instructions, total number of
edges, in-degree, and out-degree.

\noindent\textbf{Softmax Aggregation}: Similar to other recent works such as
HermesSim \cite{he2024code}, we use the Softmax aggregation function as defined
as follows:

\begin{equation}
  \mathit{softmax}(X|t) = \sum_{x_i \in X} \frac{\mathit{exp}(t \cdot x_i)}{\sum_{x_j \in X} \mathit{exp}(t \cdot x_j)} \cdot x_i
\end{equation}

Where \(t\) controls the softness of the softmax aggregation over the input
feature set \(X\), which we set to 1. This differs from the usage
within \cite{he2024code} in two key ways. Firstly, we adopt the vanilla version
of softmax aggregation which omits the multi-headed attention extensions
introduced in \cite{he2024code}. And secondly, we use the softmax aggregation
function as the GNN's aggregation function as opposed to using it as a global
graph pooling operator. The softmax operator is used to create node-level
representations at each message passing step.

\noindent\textbf{Layer-Wise Connections + Max Pool}: We opt to have Layer-Wise
feature connections from each of the model layers to store the node level
representations from each convolution. The node-level representations are stored
from each layer-wise connection before each of them being max pooled to create a
graph level representation where the resulting feature vector is a feature-wise
maximum across the node features of that layer. This operation can be seen in
Equation~\ref{eq:instance}.

\begin{equation}
  X_{i} = max_{n=1}^{N_i}X_n
\end{equation}
\label{eq:max_pool}

Global max pooling was chosen to provide a means of identifying the most
distinct features from across the call graphlet node representations. This was
viewed favorably given our objective is to identify similar functions across
architectures, compilers and optimization levels which requires distinct
representations.

\noindent\textbf{Mapping to Embedding}

The max-pooled graph-level embeddings are concatenated as:
\[X_c = [X_1 \parallel X_2 \parallel X_3] \in \mathbb{R}^{3d_{in}}\]

Where $\parallel$ denotes concatenation and $d_{in}$ is the intermediate hidden
dimension. This feature vector is then transformed through a two-layer linear
network:
\[h_1 = f_1(X_c) \in \mathbb{R}^{3d_{in}}\]
\[h_2 = f_2(h_1) \in \mathbb{R}^{3d_{in}/2}\] The final embedding is obtained
through L2 normalization:
\[z = \frac{h_2}{|h_2|_2}\]

\subsection{Training}

The model was trained using pair-based deep-metric learning alongside a
BatchHard mining scheme \cite{hermans2017defense}. We paid special attention to
the sampling methodology to ensure that each batch contained multiple (2 in our
case) versions of any function present. This ensured that it was always possible
to create a positive pair for each function present as well as provide a rich
range of possible negative samples.

The model's configuration was a hidden dimension size of 256 and an output
embedding dimension of 128. Circle loss \cite{sun2020circle} was used as the
loss function configured with $m$ set to 0.25 and $\gamma$ set to 256. Circle
loss and the hyperparameters were chosen due to good results presented in
previous works \cite{collyer2023faser}. The model was optimized using the Adam
optimizer \cite{kingma2014adam}. The model was trained for 350 epochs with each
epoch having 100K graphs randomly sampled from the total training dataset. A
cosine simulated annealing with warm restarts learning rate schedule starting at
0.0005 and finishing at 0.0001 was used, and each batch had 256 graphs within
it. The number of epochs before warm restart started at 50 and doubled for 2
subsequent times (50 -> 100 -> 200) for a total of 350 training epochs.

\section{Evaluation}

We implement \textit{KYN} using PyTorch Geometric \cite{fey2019pygeometric}. We
use \textit{bin2ml}\cite{bin2ml} alongside \textit{radare2}\cite{radare2} to generate the
data for all the experiments outlined below. All experimentation was conducted
on a workstation with 64GB RAM alongside a NVIDIA 4090 24GB GPU. 

We conducted several in-depth experiments to answer the following research
questions:\\
\noindent\textbf{RQ1:} How does \textit{KYN} perform when faced with unseen
libraries, compilers and compiler options in a cross-architecture setting?\\
\noindent\textbf{RQ2:} How does \textit{KYN} perform when faced with unseen
libraries, compilers and compiler options in a mono-architecture setting?\\
\noindent\textbf{RQ3:} How does \textit{KYN} perform when faced with function
inlining when this is not present in the training data?\\
\noindent \textbf{RQ4:} Is \textit{KYN} capable of performing zero-shot
architecture search? \\
\noindent\textbf{RQ5:} What aspects of the \textit{KYN} model contribute to it's
overall performance?

\subsection{Tasks}

The evaluation tasks have been formulated to provide a means of holistically
evaluating the performance of KYN over a large collection of binaries drawn from
several benchmark datasets. 

For \textit{RQ1}, \textit{RQ2} and \textit{RQ3} we adopt the same approach
described in \cite{marcelli2022machine} by creating search pools of \(N\) size
whereby for a given function \(f\), the generated search pool includes a single
positive and \(N-1\) negatives. Due to reporting both cross-architecture and
mono-architecture results, we have two key subtasks:

\begin{enumerate}
  \item For cross-architecture datasets, we report results for the \textit{XM}
  task described in \cite{marcelli2022machine}. The \textit{XM} task imposes no
  restrictions on what architecture, bitness, compiler or optimization level can
  be chosen and can be considered representative of real life. 
  \item For mono-architecture datasets (which are all x86-64), we report results
  comparable to the \textit{XC} task described in \cite{marcelli2022machine}.
  The \textit{XC} task imposes restrictions on the architecture and bitness that
  can be chosen but no restrictions on compiler or optimization level.
\end{enumerate}

For both cross and mono-architecture evaluation, we provide results across a
range of search pool sizes ranging from 100 to 10,000. The search pool size of
100 and 10,000 are used to compare against other SOTA approaches with the
remaining provided for future researchers to compare against. 

For \textit{RQ4}, we adopt the vulnerability search task described in
\cite{marcelli2022machine} with an addition of compiling the same version of
\textit{libcrypto.so} for RISC-V 32-bit and PowerPC 32-bit. And finally for
\textit{RQ5}, we evaluate the architecture choice impact using the \textit{XM}
tasks described above.

\begin{table*}[!ht]
  \begin{center}
  \begin{tabular}{@{}llrrrr@{}} \cmidrule(lr){3-4} \cmidrule(lr){5-6} & &
    \multicolumn{2}{c}{100}                              &
    \multicolumn{2}{c}{10,000}                           \\ \cmidrule(l){1-6}
    {Approach}                  &   {Summary}                        &
    \multicolumn{1}{c}{R@1} & \multicolumn{1}{c}{MRR@10} &
    \multicolumn{1}{c}{R@1} & \multicolumn{1}{c}{MRR@10} \\ \cmidrule(lr){1-6}
    Trex \cite{pei2020trex} & Disasm + Traces & 0.24 & 0.34 & 0.09 & 0.11 \\
    GMN \cite{li2019graph}                & CFG + BoW opc 200         & 0.45 &
    0.54                       & 0.11                    & 0.16 \\
  FASER NRM \cite{collyer2023faser}    & ESIL IR Strings           & 0.51 & 0.5
  & -                       & -                          \\
  GraphMoco \cite{sun2024graphmoco}        & ASM Basic Block + GNN     & 0.55 &
  0.60 & -                       & -                          \\
  BinFinder \cite{qasem2023binary} & VEX IR + Metadata         & 0.73 & 0.80 & -
  & -                          \\
  HermesSim \cite{he2024code}         & PCode SOG & \textbf{0.75} &
  \textbf{0.80}              & 0.44                    & \textbf{0.51} \\
  KYN (Ours)        & Call Graphlets + Metadata & \textbf{0.75} & 0.79 &
  \textbf{0.45}           & 0.50 \\
  \cmidrule(r){1-6} 
  \end{tabular}
  \end{center}
  \caption{Dataset-1 (Cisco Talos Binary Similarity \cite{marcelli2022machine})
  Test Results across search pools of 100 and 10,000 compared against relevant
  benchmarks}
  \label{ret:rq1-dataset1}
\end{table*}

\subsection{Metrics}

To facilitate robust comparison, we adopt the same metrics used within the
domain of Recall at 1 (R@1) and Mean Reciprocal Rank at 10 (MRR@10)
\cite{marcelli2022machine}. These metrics are used for all results presented as
part of RQ1, RQ2 and RQ3. At points throughout reporting the results for RQ1,
RQ2 and RQ3 we also report Normalized Discount Cumulative Gain at 10 (NDCG@10).
The NDCG@10 scores have been provided to aid future researchers compare against
our approach. For RQ4, we report the ranks at which the target vulnerable
function was present at after the search was conducted. This is the same as
previous studies \cite{marcelli2022machine}. We also report the mean and median
rank to support comparison similarly to Collyer\cite{collyer2023faser}.

\subsection{Baseline Approaches}
In order to robustly evaluate our approach, we have chosen a range of
state-of-the-art baselines approaches which include both graph and sequence
based models.

\begin{enumerate}
  \item \textbf{Trex} \cite{pei2020trex} is a transformer model trained on both
  disassembly and micro traces in order to learn execution semantics before then
  being fine-tuned for the binary function search task.
  \item \textbf{GMN} \cite{li2019graph} is a one of the GNN methods proposed in \cite{li2019graph} and has
   been shown to outperform alternative baseline graph neural
  network approaches \cite{marcelli2022machine}. This approach does however rely
  on 1-to-1 matching across the entire comparison corpus which is a significant
  drawback. 
  \item \textbf{FASER} \cite{collyer2023faser} is cross-architecture binary code
  similarity detection approach that combines long context transformers with
  radare2's native intermediate representation.
  \item \textbf{GraphMoco} \cite{sun2024graphmoco} is a hybrid model which first
  encoded basic blocks before then generating embedding representations with a
  GNN. 
  \item \textbf{Binfinder} \cite{qasem2023binary} is a simple feed-forward
  neural network which leverages VEX IR and several statistical features to
  identify similar features.
  \item \textbf{HermesSim} \cite{he2024code} is a state-of-the-art approach that
  leverages a novel graph data structure named \textit{Semantic Orientated
  Graph} alongside a graph neural network to embed functions before then
  comparing them.
\end{enumerate}

For the mono-architecture experiments, we also compare against \textbf{jTrans}
\cite{wang2022jtrans}. jTrans is a transformer model trained using a novel jump- orientated 
pre-training task on a large x86-64 dataset and can be considered as
one of the state-of-the-art models for x86-64 mono-architecture binary function
search.

\subsection{Cross Architecture Binary Function Search (RQ1)}

\subsubsection{Dataset 1}

\begin{table*}[!ht]
  \begin{center}
  \begin{tabular}{@{}llrrll@{}} \cmidrule(lr){3-4} \cmidrule(lr){5-6} & &
    \multicolumn{2}{c}{XA + XO}                             \\ \cmidrule(l){1-6}
    {Approach}                  &   {Summary}               &
    \multicolumn{1}{c}{R@1} & \multicolumn{1}{c}{MRR@10}   \\ \cmidrule(lr){1-6}
    Trex \cite{pei2020trex} & 512 Tokens                & 0.46 & 0.53 \\
  GNN \cite{li2019graph}    & CFG + BoW opc 200         & 0.57 & 0.67 \\
  GMN  \cite{li2019graph}   & CFG + BoW opc 200         & 0.61 & 0.71 \\
  KYN (Ours)                 & Call Graphlets  & \textbf{0.75} & \textbf{0.81}
  \\ \cmidrule(r){1-6} 
  \end{tabular}
  \caption{Dataset-2 Test Results across search pools of 100 compared against
  relevant benchmarks. Comparison scores taken directly from
  \cite{marcelli2022machine}}
  \label{ret:dataset2}
\end{center}
\end{table*}

The results presented in the Table~\ref{ret:rq1-dataset1} show the results
generated for search pools 100 and 10,000. Starting first with the results for
the smaller, 100 search pool size, our approach performs comparably to the
current state-of-the-art approaches, HermesSim \cite{he2024code} and BinFinder
\cite{qasem2023binary} with reported metrics that are near identical. When
comparing against our other chosen SOTA GNN baselines, GraphMoco
\cite{sun2024graphmoco} and GMN \cite{li2019graph}, our approach significantly
outperform them across all collected metrics with a R@1 relative increase of
27\% and 42\% and a MRR@10 relative increase of 24\% and 32\% respectively. When
compared against two state of the art cross-architecture Transformer models
FASER \cite{collyer2023faser} and Trex \cite{pei2020trex}, our approach again
significantly outperforms them with a R@1 relative increase of 32\% and 68\% and
a MRR@10 relative increase of 37\% and 57\%. Turning now to the larger search
pool size of 10,000, our proposed approach again reports near identical results
to HermesSim with a marginally better R@1 but a lower MRR@10 suggesting both
approaches are performing similarly.

In addition to the search pools sizes of 100 and 10,000, we also present results
for the search pool sizes of 250 and 1,000 in Table~\ref{ret:rq1-dataset1-more-sps}. This was primarily due to how significant the
difference is for the results between 100 and 10,000 is and provides a means of
reasoning about the performance degradation as the pool size increases as well
as providing future researchers empirical results to compare against.

The results presented in Table~\ref{ret:rq1-dataset1-more-sps} show that, as
expected, all metrics degrade as the search pool size increases, with the
degradation occurring steadily. Notably however, our approach outperforms the
bottom four baselines (Trex, GMN, FASER, GraphMoco) when faced with search pools
containing 1,000 functions compared to when each of the approaches is dealing
with only 100 functions. This demonstrates the effectiveness of our approach,
especially when faced with a larger number of possible candidate functions to
search.

\subsubsection{Dataset-2}

The results shown in Table~\ref{ret:dataset2} demonstrate that our approach
significantly outperforms all the baseline approaches on \textit{Dataset-2}.
Compared to the best baseline, our method achieves a 19\% increase in R@1 and a
12\% increase in MRR@10. Interestingly, the performance of our approach is
comparable to the results we report for Dataset-1. This is not the case for the
baseline approaches, which perform better on this dataset than on Dataset-1.
This suggests that not only does our model outperform the baselines in terms of
raw scores, but it also exhibits more consistent performance across different
datasets which suggests enhanced generalizability.

\begin{table}[H]
  \begin{center}
  \begin{tabular}{lrrr}
  \toprule
    {Search Pool Size} & {R@1} & {MRR@10} & {NDCG@10} \\ \cmidrule(l){1-4} 250 &
    0.664 & 0.703 &  0.726 \\
    1,000 &  0.554 & 0.611 &  0.638 \\
  \bottomrule
  \end{tabular}
  \end{center}
  \caption{Results for Dataset-1 Across Different Search Pools}
  \label{ret:rq1-dataset1-more-sps}
\end{table}

Similarly to Dataset-1, we also generated results for search pools of 250 and
1,000. The results for this are presented in Table~\ref{ret:dataset2table5}. As
the search pool size increases, the performance degradation is more significant.
We think this is likely due to the composition
of the Dataset-1 test set. The test set split is dominated by the \textit{Z3}
library (in terms of on-disk size and number of functions) resulting in a test
sample with limited variety. When faced with a dataset with more scope for
similar functions from different binaries and when at large pool sizes, the task
becomes more challenging therefore reducing overall performance.

\begin{table}[H]
  \begin{center}
  \begin{tabular}{lrrr}
  \toprule
    {Search Pool Size} & {R@1} & {MRR@10} & {NDCG@10} \\ \cmidrule(l){1-4} 250 &
    0.639 & 0.716 &  0.753 \\
    1,000 & 0.536 & 0.619 &  0.660 \\
  \bottomrule
  \end{tabular}
  \end{center}
  \smallskip
  \caption{Results for Dataset-2 Across Different Search Pools}
  \label{ret:dataset2table5}
\end{table}

\subsubsection{Binkit Noinline}

The results from evaluating our approach for \textit{binkit-noinline}
architecture presented in Table~\ref{ret:binkitnormalrq1}. Our approach again
performs consistently across all the metrics collected and reports similar
results to those collected when using \textit{Dataset-1} and \textit{Dataset-2}. 

\begin{table}[H]
  \begin{center}
  \begin{tabular}{lrrrr}
  \toprule
    {Search Pool Size} & {R@1} & {MRR@10} & {NDCG@10} \\ \cmidrule(l){1-4} 100 &
    0.828 & 0.874 & 0.895  \\
    250   & 0.777 & 0.834 & 0.860  \\
    1,000  & 0.663 & 0.728 & 0.759  \\
    10,000 & 0.442 & 0.534 & 0.582  \\
  \bottomrule
  \end{tabular}
  \end{center}
  \smallskip
  \caption{Binkit-Noinline Metrics Across Different Search Pools}
  \label{ret:binkitnormalrq1}
\end{table}

\subsubsection{Summary}

The results above answer \textit{RQ1} and demonstrate that, after robust
evaluation across a range of different binary function search benchmark
datasets, that our approach performs equally or significantly better when
compared against baseline approaches. The breadth of datasets used for
evaluation also demonstrates that our approach could be considered a general
approach which can generalize to unseen binaries, compiler options and
compilers.

\subsection{Mono-Architecture Binary Function Search (x86-64) (RQ2)}

Turning now to our mono-architecture evaluation. For this, we use two distinct
KYN models, one trained in a cross-architecture fashion (the same as the one
used to generate the results above) and another trained solely with x86-64
examples. 

The results from these experiments are in Table~\ref{ret:monoarchd1}. Both of
our model variants out-perform the baseline approaches across both of the
reported search pool sizes of 100 and 10,000. Our KYN-Cross model performs the
strongest within the 100 search pool size with a relative percentage increase of
19\% R@1 and 11\% MRR@10 when compared against jTrans and a relative percentage
increase of 6\% R@1 and 3\% MRR@10 when compared against HermesSim. The
difference in performance increases significantly when looking at the large
search pool size of 10,000. Both model variants significantly outperform both of
the benchmark approaches with the relative percentage increases of 52\%
(KYN-Cross) / 53\% (KYN-x8664) R@1 and 38\% (KYN-Cross) / 37\% (KYN-x8664)
MRR@10 when compared against jTrans. When compared against HermesSim, the
performance increase is less but still significant with a relative percentage
increase of 12\% (KYN-Cross) / 13\% (KYN-x8664) R@1 and 9\% (KYN-Cross) / 7\%
(KYN-x8864). 

\begin{table}[H]
  \begin{center}
  \begin{tabular}{lll}
    \toprule
    \textbf{}   & \multicolumn{2}{c}{Cisco D1 - XC} \\
    & 100 & 10,000 \\
    \midrule
    jTrans & 0.65/0.738 & 0.314/0.374 \\
    HermesSim & 0.756/0.807 & 0.481/0.546 \\
    \midrule
    KYN-Cross & 0.802/0.828 & 0.546/0.599 \\ 
    KYN-x8664 & 0.772/0.804 & 0.554/0.591 \\
    \bottomrule
  \end{tabular}
  \end{center}
  \caption{x86-64 Mono-Architecture results on Cisco Dataset-1. Results are presented with R@1 first and then MRR@10.}
  \label{ret:monoarchd1}
\end{table}  

To further evaluate our model and understand its performance when faced with a
larger x86-64 dataset, we also evaluated it against BinaryCorp-3M with and
without duplicates. These results can be seen in Table~\ref{ret:monobinarycorp}
and show that both model variants perform similarly with the cross-architecture
variant typically edging ahead. There are several areas of note. Firstly, the
performance of the models when faced with a different mono-architecture dataset
is significantly degraded. This highlights the need to evaluate against a range
of different dataset to get a true understanding of a models performance.
Secondly, the performance is inflated when duplicates are present across both
search pool sizes tested. This suggests that the inclusion of duplicates within
evaluation datasets is a critical mistake and may lead to misrepresentation of
the quality and performance of approaches. 

\begin{table}[h]
  \begin{center}
  \resizebox{\columnwidth}{!}{
  \begin{tabular}{lllll}
  \toprule
  \textbf{}  & \multicolumn{2}{c}{BinaryCorp3M-WD} &
  \multicolumn{2}{c}{BinaryCorp3M-ND} \\
                  & 100       & 10,000     & 100       & 10,000     \\ 
 \midrule 
  KYN-Cross   & 0.672/0.735 & 0.388/0.441 & 0.624/0.694 & 0.312/0.369 \\
  KYN-x86-64  & 0.67/0.756 & 0.366/0.419 & 0.518/0.58  & 0.314/0.362 \\ 
  \bottomrule
  \end{tabular}
  }
  \smallskip
  \caption{Mono-Architecture Binary Function Search Results. R@1 and MMR@10. WD
  denotes "with duplicates" whilst ND denotes "no duplicates"}
  \label{ret:monobinarycorp}
\end{center}
\end{table}

The evaluation results show that the two KYN model variants (KYN-Cross and
KYN-x8664) outperform the baseline jTrans and HermesSim approaches across
different search pool sizes for the mono-architecture setting, with relative
performance increases ranging from a few percent to over 50\% depending on the
metric and search pool size. Additionally, the presence of duplicates in the
evaluation datasets is shown to significantly inflate the performance metrics,
suggesting that including duplicates can misrepresent the true quality and
performance of the approaches. 

\subsubsection{Summary}

The results above answer \textit{RQ2} and demonstrate that our KYN architecture
model outperforms the baseline approaches significantly when faced with unseen
libraries, optimization options and compilers within a mono-architecture
setting. We also conduct an additional experiment that highlights the impact
duplicate functions can have on evaluation performance and hope that these
findings will be used to enhance future research methodologies.

\begin{table*}[htb]
  \begin{center}
    \begin{tabular}{llrrlrr}
    \toprule
    & ARM32 & Mean Rank & Median Rank & MIPS32 & Mean Rank & Median Rank \\ 
    \midrule
    ARM32 & 1, 1, 1, 1  & 1 & 1 & 68, 14, 85, 182, 4, 44 & 66 & 56 \\
    MIPS32 & 8, 8, 347, 1 & 91 & 8 & 2, 2, 301, 12, 1, 1 & 53 & 2 \\
    X86-64 & 1, 1, 1, 1 & 1 & 1 & 79, 14, 125, 119, 6, 72 & 69 & 75.5 \\
    X86-32 & 1, 1, 1, 1 & 1 & 1 & 74, 14, 55, 144, 22, 33 & 57 & 44 \\
    \midrule
    RISCV-32 & 1, 1, 1, 326 & 82 & 1 & 106, 17, 34, 59, 3, 17 & 39 & 25.5 \\
    PowerPC32 & 967, 192, 76, 2 & 309 & 134 & 491, 467, 583, 63, 991, 3 & 433 &
    479 \\
    \bottomrule  
    \end{tabular}
    \smallskip
    \caption{Zero Shot Vulnerability Search Results}
    \label{ret:zeroshotrq4}
  \end{center}
  \end{table*}

\subsection{Cross-Architecture Binary Function Search with Inlining (RQ3)}

The following section evaluates KYN in the presence of the potential of function
inlining. Function inlining introductions the possibility for callee functions
to be merged into (or inlined) into a caller function. If this occurs, the
structure of the caller function can change significantly.  This is an area
which is typically highlighted as a limitation of approaches or something that
is explicitly turned off \cite{qasem2023binary, marcelli2022machine}. We chose
instead of tackle this area head-on and evaluate our approach against a
\textit{BinKit} variant called \textit{BinKit-Normal}. Within this dataset, all
the binaries have been compiled where there is a possibility for function
inlining. The results presented below in Table~\ref{ret:cross-inlinerq3} show
the metric value for a given search pool size as well as the percentage
difference when compared against the results present for \textit{BinkitNoInline}
in Table~\ref{ret:binkitnormalrq1}.

\begin{table}[H]
  \resizebox{\columnwidth}{!}{%
  \begin{tabular}{lllllll}
  \toprule
    {SP} & {R@1} & {$\Downarrow$} & {MRR@10} & {$\Downarrow$} & {NDCG@10} &
    {$\Downarrow$} \\ \cmidrule(l){1-7} 100   & 0.722 & 14.7\% & 0.781 & 11.9\%
    & 0.809 & 10.6\% \\
    250   & 0.648 & 16.6\% & 0.716 & 14.1\% & 0.749 & 12.9\% \\
    1,000  & 0.569 & 14.8\% & 0.620 & 14.8\% & 0.648 & 14.6\% \\
    10,000 & 0.317 & 28.3\% & 0.402 & 24.7\% & 0.456 & 21.7\% \\
  \bottomrule
  \smallskip
  \end{tabular}
  } \caption{Binkit-Normal Metrics Across Different Search Pools with comparison
  against Binkit-NoInline. The percentages represent percentage decrease when
  function inlining was present.}
  \label{ret:cross-inlinerq3}
\end{table}

The key observations from the results are as follows. Across all search pool
sizes, there is a performance degradation of between 10\% to 28.3\% with an
average performance degradation of 16.6\%. These results do show that function
inlining does have a negative impact on our method's performance but is not
dramatic. Considering that our approach was trained with a dataset that does not
include function inlining at all, the results show our model has learned general
representation allowing it to maintain reasonable performance when faced with
function inlining. This is something that is typically not explored within other
works.

\subsubsection{Summary}

The results above answer \textit{RQ3} and show that our model is fairly robust
to function inlining even though it was trained on a dataset which does not
include inlined examples. We attribute this to our novel call graphlet input
data structure as well as the models' architecture which enable the model to
learn general representations of function neighborhoods.

\subsection{Zero-Shot Cross-Architecture Vulnerability Search (RQ4)}

We also conduct a zero-shot experiment using the same methodology as the N-day
vulnerability search task described in \cite{marcelli2022machine} but compile
the same \textit{libcrypto.so} library for RISC-V and PowerPC 32-bit to evaluate
our approaches model within a zero-shot context with the results reported in
Table~\ref{ret:zeroshotrq4}.

The performance within a zero-shot context are mixed depending on the
architecture used for search pool functions (i.e the architecture of what is
being searched). Focusing on the ARM32 firmware image, KYN demonstrates good
performance in identifying OpenSSL vulnerabilities when provided with ARM32,
x86-64, and x86-32 query functions, evidenced by a mean rank of 1 and median
rank of 1. However, when faced with a MIPS32 query function, the model's
performance degrades, with a mean rank of 91 and a median rank of 8. Regarding
zero-shot architectures, the RISCV-32 performance is acceptable, with a mean
rank of 82 and a median rank of 1, albeit skewed by a single poor retrieval.
Conversely, the PowerPC results are significantly worse, with a mean rank of
309, a median rank of 134, and a particularly high ranking of 967, indicating
high variance in the model's performance depending on the architecture
encountered. 

In addition to the ARM32 firmware image, we also report the results for the
MIPS32 image within the same dataset. This is something that other works have
typically omitted. Examining the MIPS32 architecture firmware, the results with
a MIPS32 query function are satisfactory but inferior to the ARM32 vs ARM32
results, with a mean score of 53 and a median of 2. When considering the
in-domain architectures of ARM32, X86-64, and X86-32, the model performs
acceptably, with mean ranks ranging from 57-69 and median ranks from 44-75.5,
albeit not as well as the ARM32 vs In-Domain experiments. Notably, for the
out-of-domain RISCV-32 architecture, the mean and median ranks outperform all
in-domain architectures. However, consistent with the ARM32 results, PowerPC32
performs poorly, with a mean rank of 433 and a median rank of 479.

\subsection{Summary}
The results above provide mixed evidence for RQ4. The results show promise when
compared against those reported within \cite{collyer2023faser}, especially for
zero-shot performance of RISCV-32 \(\to\) ARM32 but do not demonstrate general
zero-shot performance across all of those architectures tested. Our findings are
however consistent with Collyer \cite{collyer2023faser} in terms of MIPS32 being
inherently more difficult and challenging that other architectures. The reasons
for this are unclear and may warrant additional research.

\subsection{Ablation Study (RQ5)}

In order to understand the performance of our model in more detail and the
impact of our design decisions, we trained several models on the same dataset
and then evaluated them on the same test set, utilizing a search pools of 100
for evaluation generate empirical evidence. \textit{KYN} is the model
architecture used within the rest of the paper above, \textit{KYN-NE} is the
same model architecture, but the input graphs have no edge weights and therefore
the edge support for \textit{GraphConv} layers was not used and \textit{KYN-NES}
further amended whereby we replace the softmax aggregation function for add
within each of the convolutional layers. The results for each of these settings
can be seen below in Table~\ref{ret:ablation}.

\begin{table}[!ht]
\begin{center}
\begin{tabular}{lrr}
  \toprule
  Setting & MRR@10 & R@1 \\
  \midrule
  KYN & 0.80 & 0.76 \\ 
  KYN-NE & 0.76 & 0.71 \\
  KYN-NES & 0.75 & 0.7 \\
  \bottomrule
\end{tabular}
\end{center}
\caption{Metric scores across search pools of size 100}
\label{ret:ablation}
\end{table}

The results demonstrate that both the softmax aggregation and edge features
contribute to increasing performance. The softmax aggregation provides a modest
performance absolute improvement of 0.01 for both MRR@10 and R@1. The edge
feature information however contributes to a significant absolute improvement of
0.04 MRR@10 and 0.05 of R@1. These results suggest that incorporating edge
weights, in our case edge-betweenness, increases the quality of the function
representations for graph neural network based binary code similarity detection
tasks. 

During the experimentation and model development for this study, several
experiments were run to understand the usefulness of various types of graph
normalization techniques. Namely, Layer Normalization \cite{ba2016layer},
\textit{GraphNorm} Normalization \cite{cai2021graphnorm} and Instance
Normalization \cite{ulyanov2016instance}. To understand the effect of each, we
trained the same model with the same train and test sets but changed the
normalization technique used within each of the 3 GNN layers. 

\begin{table}[!ht]
  \begin{center}
  \begin{tabular}{lrr}
    \toprule
    & MRR@10 & R@1 \\
    \midrule
    KYN + GraphNorm & 0.52 & 0.46 \\
    KYN + LayerNorm & 0.63 & 0.57  \\
    KYN + InstanceNorm & 0.80 & 0.76 \\    
    \bottomrule
  \end{tabular}
\end{center}
\caption{Metric cross across Normalization Choices}
\label{ret:ablationnorm}
\end{table}

The results presented for the normalization experiments in Table~\ref{ret:ablationnorm} 
are clear. They suggest that, for our training data at
least, the chosen normalization approach has a profound impact of the overall
performance. When comparing GraphNorm to Instance Normalization, the impact of
choosing GraphNorm over Instance Normalization is approximately a 0.3 absolute
difference or a 42\% relative difference. The effects are marginally less for
LayerNorm but still significant with approximately a 0.17 absolute difference or
25\% relative difference when compared against InstanceNorm. 

The reason for this large performance difference likely lies within the large
variations present within our data, in particular with the number of
instructions within our chosen coarse node feature vectors and how both
LayerNorm and GraphNorm apply normalization. Both of these normalization
approaches seek to learn a normalization function for batches of graphs. Given
that each of our batches contain graphs sampled from across both architectures
and software projects, the variation within the batches is likely to be high.
During training, the normalization functions learnable parameters are optimized
to reduce this variance leading to improved model performance. When moving to the test set, 
the feature distribution changes due to being faced with software binaries which are out of distribution.
This means that when the normalization is applied, it leads to poor performance. This is especially true when you
look at the binaries contained within the train and test sets of the
\textit{Dataset-1} where the model is initially faced with several modestly
sized projects during training before then being faced with a test dataset
dominated by a very large and complex project (z3). This issue however does not
affect Instance Norm because the normalization is applied to each graph
independently within a batch. 

\subsection{Summary}

In terms of answering RQ5, the combination of Instance Normalisation, Softmax
Aggregation and edge weights all contribute to the performance of our approach.
Our first ablation study clearly show that the inclusion of edge weights
provides the biggest performance increase with a modest increase provided by the
softmax aggregation. The findings from our second ablation study show a dramatic
impact of normalization choice which we attribute to the feature variance across
different architectures and software libraries. 

\section{Limitations}
Despite the promising results achieved by the KYN approach, there are several
limitations that warrant further investigation. One potential limitation lies in
the computational requirements associated with generating the call graphlet
representations. As the size and complexity of the target binaries increase, the
computational overhead involved in extracting call graphs and constructing the
graphlet neighborhoods may become prohibitive, especially in
resource-constrained environments or time-sensitive applications. Future work
could explore more efficient techniques for graph extraction and representation.

Another limitation of the current approach is its reliance on accurate
disassembly and call graph recovery. While the methodology employed in this work
assumes reliable disassembly and call graph extraction, real-world scenarios may
involve obfuscated or heavily optimized binaries, which could impede accurate
recovery of these structures. Techniques such as control flow flattening,
virtualization obfuscation and extreme inlining could potentially undermine the
effectiveness of the proposed approach. To address this limitation, future
research will investigate incorporating deobfuscation strategies and developing
more robust graph recovery mechanisms that can handle adversarial obfuscation
techniques.

The current evaluation framework primarily focuses on open-source
software and benchmark datasets. While these provide a valuable
foundation for assessing the approach's performance, they may not fully capture
the diversity and complexity present in real-world binary analysis scenarios,
such as those encountered in embedded software reverse engineering or malware
analysis settings. Future work should aim to expand the evaluation scope by
incorporating a broader range of proprietary software, firmware, and malware
samples to further validate the approach's generalizability and robustness under
diverse and potentially adversarial conditions.

\section{Conclusions}
In this work, we proposed a novel graph neural network architecture called KYN
that leverages one-hop call graphlets for effective cross-architecture binary
function similarity detection. Our approach demonstrated state-of-the-art
performance across multiple evaluation tasks spanning several benchmark
datasets, consistently outperforming previous methods in both cross-architecture
and mono-architecture settings. Furthermore, KYN exhibited promising results
even in challenging scenarios involving function inlining and zero-shot
vulnerability search across unseen architectures. Through ablation studies, we
highlighted the importance of incorporating edge features and leveraging
instance normalization in our model design. Overall, the proposed KYN approach
provides a robust, zero-shot capable and generalizable solution for binary code
similarity detection, with potential applications in areas such as N-day
vulnerability detection, software forensics and malware analysis. Future
research directions include extending the approach to handle advanced
obfuscation techniques and exploring its applicability to other domains within
the binary analysis landscape.

\ifCLASSOPTIONcompsoc
  \section*{Acknowledgments}
\else
  \section*{Acknowledgment}
\fi

I would like to acknowledge the significant contribution of the Marcelli et al
\cite{marcelli2022machine} who have provided a foundation for robust comparison
within this domain and stimulated significant research interest.



%

\bibliographystyle{IEEEtran}
\bibliography{IEEEabrv, main.bib}

\end{document}